\documentclass[]{elsart}
\usepackage{graphicx, amsmath, amsfonts, amssymb, bm}
\usepackage[]{psfrag}
\begin{document}
\runauthor{Macovei, Xie}
\begin{frontmatter}
\title{Two-photon cooling of a nonlinear quantum oscillator}
\author{Mihai A. Macovei} 
\footnote{On leave from Institute of Applied Physics, 
Chi\c{s}in\u{a}u, Moldova},
\ead{mihai.macovei@mpi-hd.mpg.de}
\author{Xiao-Tao Xie}
\address{Max-Planck-Institut f\"ur Kernphysik, Saupfercheckweg 1, D-69117 Heidelberg, Germany}
\date{\today}
\begin{abstract}
The cooling effects of a nonlinear quantum oscillator via its
interaction with an artificial atom (qubit) are investigated. The
quantum dissipations through the environmental reservoir of the
nonlinear oscillator are included, taking into account the
nonlinearity of the qubit-oscillator interaction. For appropriate
bath temperatures and the resonator's quality factors, we
demonstrate effective cooling below the thermal background. As the
photon coherence functions behave differently for even and odd
photon number states, we describe a mechanism distinguishing those
states. The analytical formalism developed is general and can be
applied to a wide range of systems.
\end{abstract}
\end{frontmatter}
\section{Introduction}
Simple models describing the main properties of various phenomena in
physics are always of particular importance. The Jaynes-Cummings
model, for instance, provides a simple description of the
interaction of matter with an electromagnetic field \cite{JC}. It consists of
a two-state particle interacting with a single quantized mode,
applicable in general to cavity quantum electrodynamics. Now, there
is an increased interest to apply such a simple model to more
complex systems like superconducting electrical quantum circuits. 
This allows us to investigate them in an analogous way as a two-level 
atom interacting with a quantized electromagnetic cavity mode \cite{tw_jc}. 
As an advantage, for instance, the strong coupling limit and enhanced 
lifetimes can be achieved in superconducting devices \cite{cqe_sec}. 
Remarkably, a Josephson qubit was designed that entangles qubit 
information \cite{moo}. Entanglement between a superconducting flux 
qubit and a superconducting quantum interference device was 
demonstrated in \cite{ent_f}, while a procedure to directly measure 
the state of an electromagnetic field inside a superconducting 
transmission line, coupled to a Cooper-pair box, was proposed in 
\cite{tr_l}, making them attractive to quantum computation processing. 
Of particular interest are the studies regarding the dephasing of 
the superconduction qubit induced by the photon noise \cite{deph}. 
Other systems refer to coupling of the superconducting qubit to a 
solid-state nanomechanical resonator. Various interesting effects 
related to the nanomechanical resonator state were demonstrated. 
In particular, squeezing of the nanomechanical resonator state 
occurs when coupling it to a Josephson quantum circuit \cite{sqnm}. 
The fidelity of a state transfer from the Josephson junction to a 
nanomechanical resonator was investigated in \cite{fid}. 
Entanglement from a nanomechanical resonator weakly coupled to a 
single Cooper-pair box \cite{ent_c}, continuous measurement of the 
energy eigenstates of a nanomechanical resonator without a 
nondemolition probe \cite{non_d} or signatures for a classical to 
quantum transition of a driven nonlinear nanomechanical resonator 
\cite{cl_q} were already discussed.

Via engineering superconducting elements as artificial atoms and
coupling them to a photon field of a resonator or to vibrational
states of a nanomechanical resonator one can demonstrate other
interesting phenomena such as single artificial atom lasing or
cooling. In particular, schemes for ground-state cooling of
mechanical resonators were proposed in \cite{m_cool}. A flux qubit
was experimentally cooled \cite{fl_cool} by using techniques somewhat
related to the well-known optical sideband cooling methods (see
e.g. Ref.~\cite{rmp} and references therein). Continuous monitoring
of Rabi oscillations in a Josephson flux qubit was reported in
\cite{r_mon} while lasing effects of a Josephson-junction charge
qubit, embedded in a superconducting resonator, was experimentally
demonstrated in \cite{l_exp}. Single-qubit lasing and cooling at the
Rabi frequency was proposed in \cite{qu_cool}, while a mechanism of
simultaneously cooling of an artificial atom and its neighboring
quantum system was analyzed in \cite{sim_cool}. In some of these
systems the nonlinear qubit-oscillator interaction was considered,
i.e. the case when the qubit exchanges simultaneously more than one
photon with the resonator mode. However, the dissipations of the
nonlinear system due to interaction of the quantized mode with the
environmental reservoir are more complex and require a further
treatment.

Thus here we report additional results regarding 
the nonlinear matter-light interactions which are general and 
applicable to a wide range of systems. To this end, we investigate 
the properties of a quantum oscillator coupled nonlinearly with a 
driven qubit through two-photon effects and damped via the thermal 
environmental reservoir, and focus on cooling phenomena of the 
oscillator's degrees of freedom. Due to a high degree of 
correlations between the particles generated in such a two-photon 
process, we consider the nonlinear damping of the generated photons. 
This allows us to describe the system by using the properties of 
su(1,1) algebra. In the steady-state we obtain a mean photon number 
well below unity and thermal limit. Consequently, an effective cooling 
mechanism via nonlinear processes is discussed. Further, we propose a 
scheme which is able to distinguish between even and odd photon number 
states corresponding to su(1,1) algebra via measuring the second-order 
photon coherence function (and/or higher-order photon correlations). 
In addition, photon statistics may show quantum features, i.e. an 
important step towards single-photon sources.

The paper is organized as follows. In Sec. 2 we introduce the system of 
interest and derive the corresponding master equation. The next section 
3 analyzes the results. Finally the summary is given in Sec. 4.

\section{The model}
Particularly, we consider a Josephson flux qubit coupled inductively 
to a slow LC oscillator. The frequency of the oscillator is much lower 
than the qubit's tunnel splitting, i.e. $\omega_{c} \ll \Delta$. The 
qubit is driven with Rabi frequencies near resonance with the oscillator 
frequency that affect the oscillator, increasing its oscillation amplitude. 
Near the symmetry point (i.e. the energy bias $\epsilon$ between the flux 
states is negligibly small) and after transformation to the qubit's 
eigenbasis, the Hamiltonian describing the systems is:
\begin{eqnarray}
H=\Delta\sigma_{z}/2+\Omega\cos{(\omega
t)}\sigma_{x}+\omega_{c}a_{+}a - g\sigma_{x}(a + a_{+}), \label{HMq}
\end{eqnarray}
where the first term describes the qubit while the second one
considers its driving by an applied AC magnetic flux with amplitude
$\Omega$ and frequency $\omega$. The last two terms describe the
oscillator with frequency $\omega_{c}=1/\sqrt{LC}$ as well as the
qubit-oscillator interaction, respectively. Here $g \approx
MI_{p}I_{c0}$, where $M$ is the mutual inductance, $I_{p}$ the
magnitude of the persistent current in the qubit, and
$I_{c0}=\sqrt{\omega_{c}/2L}$ the amplitude of the vacuum
fluctuations of the current in the LC oscillator. $a_{+}$ and $a$
are the creation and annihilation operators corresponding to the
oscillator degrees of freedom, while $\sigma_{i}$ $(i \in
\{x,y,z\})$ are the Pauli matrices operating in the dressed flux
basis of the qubit subsystem. As $\Delta \gg \omega_{c}$, the 
transverse coupling in the Hamiltonian (\ref{HMq}) is transformed 
into a second-order longitudinal coupling by employing a 
Schrieffer-Wolff type transformation, i.e. 
$U_{S}=\exp{(iS)}$ with $S=(g/\Delta)(a+a_{+})\sigma_{y}$
\cite{qu_cool,KSS}. By further using  the rotating wave approximation
with respect to $\omega$ and diagonalizing the qubit term as well as
applying the secular approximation, i.e. omitting terms oscillating
with the generalized Rabi frequency, one arrives at the following
Hamiltonian describing the nonlinear interaction between the qubit
and the oscillator:
\begin{eqnarray}
H&=&\Omega_{R}\sigma_{z}/2+ \omega_{c}a_{+}a +
g_{2}(a^{2}_{+}\sigma^{-} + \sigma^{+}a^{2})/2 \nonumber \\
&-& g_{0}(aa_{+} + a_{+}a)\sigma_{z}/4. \label{HM}
\end{eqnarray}
Here $g_{2}$=$2g^{2}\sin{2\theta}/\Delta$ gives the nonlinear 
qubit-oscillator coupling strength while $g_{0}$=$4g^{2}\cos{2\theta}/\Delta$ 
accounts for a frequency shift of the qubit's frequency. 
Further $\cot{2\theta}=\delta \omega/\Omega$, where 
$\delta \omega =\Delta-\omega$ and where
$\Omega_{R}=\sqrt{(\delta \omega)^{2}+\Omega^{2}}$ stands for the
generalized Rabi frequency. The Hamiltonian (\ref{HM}) involves
two-quantum processes, i.e. two-particle exchanges between the qubit
and the nonlinear oscillator, which means that the quanta are created 
and annihilated simultaneously in pairs. The particles generated via 
such a quadratic process are known to be highly correlated, i.e. a 
single photon pair behaves like a quasiparticle \cite{emm}.

The spontaneous emission damping of the qubit in this picture is given 
as \cite{qu_cool,mk}:
\begin{eqnarray}
\dot
\rho_{sp}=&-&\gamma^{(0)}[\sigma_{z},\sigma_{z}\rho]-\sum_{\alpha_{1}\not=\alpha_{2}
\in
\{+,-\}}\gamma^{(\alpha_{1})}[\sigma^{\alpha_{1}},\sigma^{\alpha_{2}}\rho]
+ {\rm H.c.}, \label{sp}
\end{eqnarray}
where $\gamma^{(+)}=\Gamma_{0}\cos^{4}{\theta}/2$,
$\gamma^{(-)}=\Gamma_{0}\sin^{4}{\theta}/2$ and
$\gamma^{(0)}=\Gamma_{0}\sin^{2}{2\theta}/8$.

The damping of the quantized oscillator mode depends on the 
environmental reservoir. In order to have a two-photon damping 
of the nonlinear oscillator we consider that the quantum oscillator 
couples with the environmental bath via the following Hamiltonian
\begin{eqnarray}
H_{f}= \hbar \nu b^{\dagger}b + 2\hbar \tilde \chi(b^{\dagger}\beta^{-}+\beta^{+}b).
\label{hf}
\end{eqnarray}
Here the operators $b^{\dagger}(b)$ belong to the broadband reservoir 
of carrier frequency $\nu$ and represent the photon generation 
(annihilation) operator for the bath. Such a reservoir can be obtained 
by assuming that the $LC$ oscillator couples additionally with another 
circuit the frequency of which $\nu$ is equal or close to 2$\omega_{c}$.
Eliminating the bath operators in the Born-Markov approximation
one can arrive at the master equation describing the damping of the 
nonlinear oscillator. For further convenience we introduce the field operators
\begin{eqnarray*}
\beta^{+}=a^{2}_{+}/2,~\beta^{-}=a^{2}/2~{\rm and}~\beta_{z}=(a_{+}a+1/2)/2
\end{eqnarray*}
which obey the commutation relations for su(1,1) algebra, i.e.
$[\beta^{+},\beta^{-}]=-2\beta_{z}$ and $[\beta_{z},\beta^{\pm}]=\pm
\beta^{\pm}$. These operators act on the corresponding bases states
of the su(1,1) algebra in the following way:
\begin{eqnarray}
\beta^{+}|j,m\rangle &=& \sqrt{(m+1)(m+2j)}|j,m+1\rangle, \nonumber \\
\beta^{-}|j,m\rangle &=& \sqrt{m(m+2j-1)}|j,m-1\rangle, \nonumber \\
\beta_{z}|j,m\rangle &=& (m+j)|j,m\rangle. \label{jm}
\end{eqnarray}
Here $m \in \{0,1,2, \cdots,\infty\}$, while for a single mode
field, as considered in our approach, the allowed value of the
Bargmann index (i.e., $j$) is $1/4$ ($3/4$) for an even (odd) photon
number. The correspondence between the number state of the single
mode field $|n\rangle$ and the su(1,1) basis states $|j,m\rangle$ is
$|n\rangle \leftrightarrow |j,m\rangle$ for $n=2(m+j)-1/2$
\cite{ger}.

We have derived the master equation corresponding to the damping of 
the nonlinear oscillator via two-photon processes which can be written 
as follows (see Appendix):
\begin{eqnarray}
\dot \rho_{f}&=&-i[H_{0},\rho]-\kappa(1+\bar n)
\bigl \{[\beta^{+},\beta^{-}\rho] + [\rho \beta^{+},\beta^{-}] \bigr \} \nonumber \\
&-&\kappa\bar n\bigl \{[\beta^{-},\beta^{+}\rho] + [\rho
\beta^{-},\beta^{+}]\bigr \}, \label{fd}
\end{eqnarray}
with $H_{0}=2\bar \chi\bar n\beta_{z}-\bar \chi\beta^{+}\beta$
describing an additional shift of the oscillator mode frequency
proportional to $\bar \chi \bar n$ and the Lamb shift proportional to
$\bar \chi$, respectively, induced by the thermostat via an effective
coupling constant $\bar \chi$. Here $\kappa$ is the two-photon damping 
rate of the quantized mode while $\bar n$ is the mean thermal photon number 
at frequency $2\omega_{c}$. In fact, for $\bar n=0$ one obtains the
well-known nonlinear damping of a quantized cavity mode via two-quantum
processes used in Cavity Quantum Electrodynamics (see for instance
\cite{ger,sczb,grkn}). Finally, the master equation characterizing our
model reads as follows:
\begin{eqnarray}
\dot \rho&=&-i[\tilde H,\rho]-\Lambda \rho_{sp}-\Lambda \rho_{f},
\label{ME}
\end{eqnarray}
where $\tilde H=2(\omega_{c}+\bar \chi\bar n)\beta_{z}-\bar \chi\beta^{+}\beta^{-}
+(\Omega_{R}-2g_{0}\beta_{z})\sigma_{z}/2 + g_{2}(\sigma^{-}\beta^{+}+\beta^{-}\sigma^{+})$, 
while $\Lambda\rho_{sp}$ and $\Lambda \rho_{f}$ are given by Eq.~(\ref{sp}) 
and Eq.~(\ref{fd}), respectively. Note here that the form of the 
master equation $(\ref{ME})$ would be similar to the corresponding 
one describing a wide range of problems involving two-quantum 
processes as for instance the quantum dynamics of a single 
two-state particle or a collection of two-state particles 
possessing dipole-forbidden transitions, pumped with an intense 
laser field in two-photon resonance and damped at resonance via 
two-photon effects by an optical cavity containing a two-photon 
absorber. Similar nonlinear damping, as in Eq.~(\ref{fd}), can be 
applied to a cavity mode crossed by an excited flux of such 
dipole-forbidden emitters. The laser/maser phenomenon via two 
photons can be developed here as well. Other applications refer to 
quantum effects in the present scheme as, for example, the first- 
and second-order squeezing of the oscillator's quantum fluctuations. 
These studies will be presented elsewhere.

\section{Results and discussions}
A general analytical solution of Eq.~(\ref{ME}) is not evident.
However, one can obtain its solution for different regimes of
interest, namely in the bad or good cavity limit. Therefore, we
proceed by investigating the properties of Eq.~(\ref{ME}) when the
qubit's quantum dynamics is faster than that of the nonlinear
quantum oscillator, i.e. in the good cavity limit. Below the photon 
saturation number $n_{0}=\bigl(\Gamma_{||}\Gamma_{\perp}/2g^{2}_{2}\bigr )^{1/2}$, 
with $\Gamma_{\perp}=4\gamma^{(0)}+\Gamma_{||}$ and
$\Gamma_{||}=\gamma^{(+)}+\gamma^{(-)}$, one can integrate the
qubit's degrees of freedom to arrive at a master equation
characterizing the quantum oscillator only:
\begin{eqnarray}
\dot \rho &-& i\bar \chi[\beta^{+}\beta^{-},\rho]=-\bigl(\kappa(1+\bar
n) + \Gamma_{-}\bigr)[\beta^{+},\beta^{-} \rho] \nonumber \\
&-& \bigl(\kappa\bar n +
\Gamma_{+}\bigr)[\beta^{-},\beta^{+}\rho] + {\rm H.c.} \label{bme}
\end{eqnarray}
Here $\Gamma_{\pm}=g^{2}_{2}(1 \pm \langle
\sigma_{z}\rangle_{0})/(2\Gamma_{\perp})$, with $\langle
\sigma_{z}\rangle_{0}=(\gamma^{(-)}-\gamma^{(+)})/\Gamma_{||}$ being
the qubit inversion in the absence of the resonator mode. The
two-photon resonance was assumed, i.e. $\Omega_{R}-2g_{0}\langle
\beta_{z}\rangle=2(\omega_{c}+\bar \chi \bar n)$, as well as the
relation: $\kappa(1 + \bar n) \ll g_{2} < \Gamma_{0}$.

The steady-state solution for the diagonal elements of
Eq.~(\ref{bme}) is
\begin{eqnarray}
\rho_{s}=Z^{-1}\exp[-\alpha \beta_{z}], \label{ss}
\end{eqnarray}
where $Z$ is determined by the requirement ${\rm Tr(\rho_{s})=1}$
and $\alpha=\ln{\eta}$, with $\eta=\bigl(\kappa(1+\bar
n) + \Gamma_{-}\bigr)/\bigl(\kappa \bar n + \Gamma_{+}\bigr)$.
The expectation values of the operators needed for evaluating the
properties of the nonlinear oscillator are obtained from
Eq.~(\ref{jm}) and Eq.~(\ref{ss}). In particular, the nonlinear
oscillator mean photon number, i.e. $\langle n\rangle=2\langle
\beta_{z}\rangle-1/2$, and its second- and fourth-order correlations
can be determined from the following expressions:
\begin{eqnarray}
\langle \beta_{z}\rangle &=& j + \frac{1}{\eta -1}, \nonumber \\
\langle \beta^{+}\beta^{-}\rangle &=& \frac{2(1+j(\eta-1))}{(\eta-1)^{2}},
\nonumber \\
\langle \beta^{+^2}\beta^{-^2}\rangle
&=&\frac{12(1+\eta)+4(\eta-1)(5+\eta)j}{(\eta-1)^{4}}+\frac{8j^{2}}{(\eta-1)^{2}}. \label{kr}
\end{eqnarray}
It can be observed here that when $\eta$ approaches unity, the result 
is a substantial increase in the photon number and photon correlations.
This will lead to lasing instability phenomena so that Eq.~(\ref{bme}) 
and its solution are not valid anymore. However, $\eta \gg 1$ 
corresponds to the cooling of the nonlinear oscillator
where the application of solution (\ref{ss}) is justified below the
photon saturation number $n_{0}$, because the mean photon number as
well as second- and fourth-order photon correlations tend to lower
values in this case. Note that the control parameter $\eta$ can be
modified by adjusting the qubit's parameters as well as the detuning
of the external driving field.
\begin{figure}[t]
\includegraphics[width=7cm]{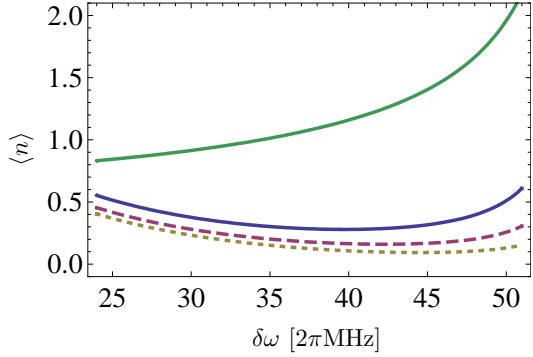}
\caption{\label{fig-1}(color online) The mean photon number of the
nonlinear oscillator $\langle n\rangle$ as a function of detuning
$\delta\omega$. The solid blue curve stands for $\bar n$=4, the
long-dashed line for $\bar n$=2, while the short-dashed one
corresponds to $\bar n$=1. The solid green curve shows the
saturation photon number $n_{0}$. Here, $\omega_{c}/2\pi$=27.5MHz,
$\kappa/2\pi$=2kHz, $\Delta/2\pi$=3GHz, $g/2\pi=$18MHz,
$\Gamma_{0}/2\pi$=0.5MHz, $\Omega =
\sqrt{\Omega^{2}_{R}-(\delta\omega)^{2}}$, and $j=1/4$.}
\end{figure}

Fig.~(\ref{fig-1}) depicts the mean photon number in the nonlinear
oscillator mode, i.e. $\langle n\rangle$ when $j=1/4$, as a 
function of various parameters governing steady-state behaviors. 
As can be observed here, lower photon numbers can be achieved via 
a suitable choice of the parameters involved and below the thermal 
limit. By increasing the coupling coefficient $g$ such that 
$g_{2} \ll \Omega_{R}$, the cooling efficiency can be further improved. 
Evidently, the qubit is more in its ground dressed-state, i.e. 
$\langle \sigma_{z}\rangle_{0}<0$ ($\delta \omega > 0$), when 
the cooling occurs. On the other hand, inversion of the qubit 
population can be created via modifying the detuning $\delta 
\omega$, that is for $\delta \omega < 0$. Thus, the cooling of 
the nonlinear oscillator occurs when controlling the qubit's 
population quantum dynamics. Although we get lower photon numbers 
for the nonlinear oscillator mode, it will be not easy, in general, 
to achieve $\langle n\rangle \approx 0$. Due to approximations 
used in our approach, we cannot increase the coupling $g$ ($g_{2}$) 
further since we have performed the rotating wave approximation 
in the Hamiltonian (\ref{HM}). The counter-rotating terms have to be 
taken into account when proceeding to larger $g$ ($g_{2}$). 
Neither in this case can the degrees of freedom related to the 
qubit's quantum dynamics be adiabatically eliminated because
$g_{2}\sim \Gamma_{0}$. Other limiting factors may appear due 
to fluctuations of external parameters. However, improving the 
oscillator quality factor one can achieve better cooling in 
general.

We focus further on the properties of photon coherences. The
second-order coherence function, i.e. $g^{(2)}(0)=4\langle
\beta^{+}\beta^{-}\rangle/\langle n\rangle^{2}$, and the
fourth-order one, i.e. $g^{(4)}(0)=\langle
\beta^{+^{2}}\beta^{-^2}\rangle/\langle
\beta^{+}\beta^{-}\rangle^{2}$, can be evaluated by using
Eq.~(\ref{jm}) and Eq.~(\ref{ss}) and represented as follows:
\begin{eqnarray}
g^{(2)}(0)&=&\frac{32\bigl(1+(\eta-1)j\bigr)}{\bigl(5+4j(\eta-1)-\eta
\bigr)^{2}}, \nonumber \\
g^{(4)}(0)&=&2 + \frac{1+3\eta+j(\eta^{2}-1)}{\bigl (1 + j(\eta-1)
\bigr )^{2}}. \label{chf}
\end{eqnarray}
An interesting result here is that the above correlation functions
behave differently for even ($j=1/4$) or odd ($j=3/4$) photon
numbers. For instance, $g^{(2)}(0)=(3+\eta)/2$ and $g^{(4)}(0)=2 +
4(3+12\eta+\eta^{2})/(3+\eta)^{2}$ when $j=1/4$, while
$g^{(2)}(0)=2(1+3\eta)/(1+\eta)^{2}$ and $g^{(4)}(0)=2 +
4(3+12\eta+3\eta^{2})/(1+3\eta)^{2}$ when $j=3/4$. Particularly, for
even or odd photon number states, the second-order coherence
function $g^{(2)}(0)$ will be linearly or inversely proportional to
$\eta$ when $\eta$ increases. Depending on the steady-state
behaviors of the photon correlation functions, one can distinguish
between the nonlinear oscillator's odd and even photon number
states. Thus, the photon coherence functions are a convenient tool
to determine the parity of the photon number of the state
$|j,m\rangle$ which corresponds to su(1,1) algebra.
\begin{figure}[t]
\includegraphics[width=7cm]{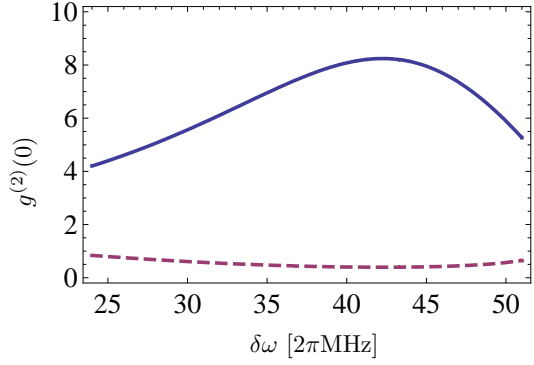}
\caption{\label{fig-2}(color online) The second-order photon
coherence function $g^{(2)}(0)$ as a function of detuning
$\delta\omega$ and for $j=1/4$ (solid line), and $j=3/4$
(long-dashed curve), respectively. The other parameters are the same
as in Fig.~(\ref{fig-1}) with $\bar n=2$.}
\end{figure}

In Fig.~(\ref{fig-2}) we show the dependence of the second-order
coherence function $g^{(2)}(0)$ versus the detuning $\delta \omega$
and different values of $j$. These behaviors are explained as
follows: for an even number of photons describing the state
$|j,m\rangle$, i.e. $j=1/4$, the mean-photon number $\langle
n\rangle$ will be below unity (for $\bar n=2$) and the normalized
second-order coherence function increases accordingly, showing
super-Poissonian photon statistics ($g^{(2)}(0)>1$). Conversely, 
for an odd number of photons, i.e. $j=3/4$, the mean-photon 
number $\langle n\rangle$ will be above unity (or near unity) and 
the second-order coherence function decreases, revealing near 
Poissonian ($g^{(2)}(0)\approx 1$) or even sub-Poissonian 
($g^{(2)}(0)<1$) photon statistics and a single-photon state 
can be created here. Note that the fourth-order coherence 
function $g^{(4)}(0)$ approximately behaves as $g^{(2)}(0)$, 
but with a different magnitude.

\section{Summary}
In summary, we described a scheme capable of cooling an
oscillator coupled to an externally pumped artificial 
atom (a Josephson flux qubit) and damped nonlinearly 
through interaction with its environmental thermal 
reservoir. Under certain conditions, the oscillator and 
the qubit exchange two-photons, allowing us to describe 
their quantum dynamics using the su(1,1) algebra. If the 
qubit's dynamics is faster than that of the nonlinear 
oscillator, the cooling of the oscillator's degrees of 
freedom occurs when controlling the qubit quantum dynamics. 
Evaluating the second-order photon correlation function 
(or higher-order correlations), one can distinguish between
even and odd photon number states characterizing the oscillator. 
By adjusting the parameters involved, one can create a nonclassical 
field state with sub-Poissonian photon statistics. This will allow 
us to obtain a single-photon state of the nonlinear oscillator.
\appendix
\section{Appendix}
In this Appendix we obtain the equation (\ref{fd}). We start by indicating  the 
Hamiltonian $H_{f}$ describing the interaction of the environmental bath with the nonlinear 
oscillator, i.e. the Eq.~(\ref{hf}):
\begin{eqnarray*}
H_{f}=\hbar \nu b^{\dagger}b + 2\hbar\tilde \chi(b^{\dagger}\beta^{-} + \beta^{+}b).
\end{eqnarray*}
In the Born-Markov approximation one can eliminate the bath operators. For doing this 
we define an operator $Q_{f}$ which belongs to the oscillator's subsystem and 
satisfy the following equation of motion:
\begin{eqnarray}
\frac{d}{dt}\langle Q_{f}\rangle = 2i\tilde \chi \{\langle b^{\dagger}[\beta^{-},Q_{f}]\rangle + 
\langle[Q_{f},\beta^{+}]b\rangle\}. \label{Qf}
\end{eqnarray}
The formal solution of the Heisenberg equation for $b^{\dagger}$ is:
\begin{eqnarray}
b^{\dagger}(t)&=&b^{\dagger}(0)e^{i(\nu + i \chi)t} - 2\tilde \chi\frac{\beta^{+}(t)}{\nu-2\omega_{c} + i\chi} \nonumber \\
&=& b^{\dagger}_{v}(t)-2\tilde \chi\frac{\beta^{+}(t)}{\nu-2\omega_{c} + i\chi}, \label{bp}
\end{eqnarray}
with $b(t)=[b^{\dagger}(t)]^{+}$. Substituting Eq.~(\ref{bp}) in Eq.~(\ref{Qf}) 
and using the Bogolubov lemma \cite{bg} 
\begin{eqnarray}
\langle b^{\dagger}_{v}(t)U(t)\rangle = -\frac{2\tilde \chi}{\nu-2\omega_{c}+i\chi}\bar n \langle[\beta^{+}(t),U(t)]\rangle,
\end{eqnarray}
where $U$ is an arbitrary operator belonging to the oscillator subsystem together with the identity 
$Tr\{\frac{d}{dt}Q_{f}(t)\rho_{f}(0)\}=Tr\{\frac{d}{dt}\rho_{f}(t)Q_{f}(0)\}$ one arrives at 
Eq.~(\ref{fd}). There $\kappa=\frac{\chi(2\tilde \chi)^{2}}{(\nu-2\omega_{c})^{2}+\chi^{2}}$ 
and $\bar \chi = \frac{(\nu-2\omega_{c})(2\tilde \chi)^{2}}{(\nu-2\omega_{c})^{2}+\chi^{2}}$.

\end{document}